\documentstyle[preprint,aps,psfig]{revtex}
\begin{document}
\draft
\preprint{LA-UR-02-5476}
\title{Hybridgen: A Model for the Study of QCD Hybrid States }
\author{M.M. Brisudova$^a$\thanks{{\it On leave from Physics Institute,
 Slovak Acad. Sci., D\'{u}bravsk\'{a} cesta 9,  842 28 Bratislava,
Slovakia.} Electronic address: {\tt brisuda@niobe.iucf.indiana.edu}},
 and
T. Goldman$^b$\thanks{Electronic address: {\tt tgoldman@lanl.gov}}
}
\address{$^a$ Nuclear Theory Center, Indiana University,
2401~Milo~B.~Sampson~Lane, Bloomington,~IN~47408\\} \address{$^b$
Theoretical Division, MS B283, Los Alamos National Laboratory, Los
Alamos,~NM~87545} 
\date{September 19, 2002}
\maketitle

\begin{abstract}

We study the mixing of excited states of a Hydrogen atom in a cavity
with de-excited states plus a confined photon as a model for the
coupling of quark-antiquark and quark-antiquark-gluon hybrid states 
in QCD. For an interesting range of parameters, the results are 
analytic. We find a case for which wavefunctions (and hence decay
patterns) may be at odds with mass with respect to identification 
of a state as hybrid or not.

\end{abstract}

\bigskip

\section{Introduction}

Hybrid (quark-gluon) states are expected to occur in QCD but have not
yet been unambiguously experimentally observed. Lattice efforts to
study hybrid states~\cite{lattice} directly are hampered by two
problems:  First, these states are generally not the lowest energy ones
of a given spin-parity combination, which reduces the statistical power
available for studying them, since the underlying states must first be
removed from the Monte Carlo signal. In addition, since, unlike
photons, individual gluon states are not gauge invariant, it is
difficult to unambiguously identify the contribution of the gluonic
component to the detailed structure and even the energy of the state.
The problems accompanying direct lattice calculations of hybrid masses
are avoided when, instead of individual physical states, static
potentials such that the glue can carry angular momentum are determined
\cite{morningstar}. Masses of physical states can then be found in the
leading Born-Oppenheimer approximation, at least for very heavy
quarks.

These lattice data~\cite{morningstar} provide a standard with which all
models may be compared (for a concise review see, e.g., \cite{philip}).
At small quark separation, lattice data are reasonably well described
by bag models~\cite{bags}. Flux tube models~\cite{swanson} form the
basis of a different physical picture of hybrids.  Hybrid states have
been studied in the Coulomb gauge Hamiltonian approach~\cite{adam},
which, owing to a nontrivial BPS vacuum, naturally leads to a mass gap.
Physically, the transverse glue in the hybrid constitutes a valence
particle.  This approach is  not applicable at large separation, where
flux tube models~\cite{swanson} and string pictures~\cite{string} seem
to provide a better description \cite{philip} in terms of oscillation
mode excitations of the string or the tube.  Earlier constituent gluon
models generally do not agree with the lattice mass
predictions~\cite{kalashnikova}.

With a view to elucidating the nature of such states, we examine a
modification of the hydrogenic problem. We describe our motivation
in choosing this analogy in detail in the next Section, providing only
an overview here.

To model the effect of gauge boson confinement, we place the hydrogen
atom in a (spherical) cavity within a magnetic superconducting material.
This prevents photon propagation away from the atom, thus confining it
to the same cavity. Separate sets of electronic and photonic eigenstates
 are thus defined. By occupying a photon state simultaneously
with an electron state, we obtain our analog hybrid state. Through
photon absorption, these states couple to the purely electronic states,
and so mix. This model also has the advantage of allowing for removal
of the degeneracy between the excited atomic state and combined lower
atomic state plus photon. This is more closely analogous to the case 
in hadronic systems, where the phenomenologically-based expectation 
is that states including gluon excitations have considerably larger
energies than corresponding states including only quarks and
antiquarks.

The paper is organized as follows. In Sec.2, we present details of
the motivation for our model. In Sec.3, we describe the electronic
and photonic states and the relevant transition matrix elements for
photon emission and absorption. In Sec.4, we discuss the energy
eigenvalues and mixing amplitudes in various limits. Finally, we
close with concluding remarks in Sec.5.

\section{Motivation: U(1) vs. SU(3)}

The definition of hybrid states is rarely made explicit unless
presented in a non-relativistic or constituent context. From a
relativistic point of view, intermediate state projections should be
able to demonstrate the contribution of gluonic excitations to the Fock
space of states, since gluon exchanges are responsible for the binding
of quarks and antiquarks in even those states not viewed as hybrid.

Such excitations are the analog of Coulomb photonic contributions to
atomic states.  In a hydrogen atom, the spectrum of states is normally
viewed as consisting entirely of electron eigenstates in the Coulomb
field of the proton (in the limit of neglecting proton recoil effects).
{}From a field theoretic point of view, however, there are cuts of the
off-shell electron-proton scattering amplitude which may traverse one
or more photons of the Coulomb exchanges producing the binding. These
are not viewed as contributions of (electron-photon) `hybrid'
intermediate states to the wavefunction since neither these photons,
nor the electrons, in the virtual three particle (including the proton)
intermediate state are in eigenstates of the overall system.

There are, nonetheless, hybrid contributions to the wavefunctions,
in principle. These can be seen for excited states, such as the 2P
state for example, when it undergoes a transition to the 1S state
and emits a photon. Although a free outgoing wave, this photon is
in an eigenstate of the overall system to leading order in the
electromagnetic perturbations. Since there is a finite, although
small, probability for this photon to be reabsorbed and promote
the 1S state back to the 2P state,\footnote{In field theoretic
terms, this constitutes a self-energy correction to the 2P state
propagator.} there is, in principle, a hybrid electron-photon
state contribution to the precise description of the nominal 2P
excited state. Part of the reason that this may generally be
ignored is that the contribution is extremely small, as the photon
is free to propagate away from the location of the atom, making
reabsorption impossible.

This is not true for the analog case in QCD. In a meson composed
of a very heavy antiquark and a light quark, for example, a higher
excited state may emit a gluon, leaving the quarks in a lower state,
except that they now occupy an overall color octet state instead of
a singlet. However, the gluon cannot propagate away, but remains
confined to the region of the color octet source formed by the
quark-antiquark pair.

Thus, to form an analogous atomic system, we must arrange for the
transition photon to be confined to the region of the parent atomic
state. We implement this by imagining the atom to be centered in a
spherical region (bubble) of vacuum contained within an effectively
infinite, perfectly conducting medium. We choose to use magnetic
superconductor boundary conditions instead of the usual (for QED)
electric superconductor boundary conditions in order to emulate the QCD
case more closely.  To avoid unnecessary additional complications, we
suppose also that the proton is fixed at the center of the cavity,
obviating concerns about both center of mass motion effects and
distortions from sphericity.

There are, however, some additional effects introduced by the
cavity that must still be accounted for:  The truncation of the 
Coulomb field of the proton (fixed at the cavity center) affects 
the electronic eigenstates by altering the boundary condition 
from the usual one at infinity, and the boundary condition 
produced by the reflecting interior surface defines a set of 
eigenmodes for the photon, replacing the continuum of outgoing 
waves previously available. This part simply corresponds to the 
well-studied case of the bag model~\cite{trn,bagschool}.

Because the quarks in the QCD case are not in a color singlet state
after the gluon emission, our analogy is imperfect. To reduce the
complications associated with this, we examine an atomic case with a
magnetic transition, which does not alter the spatial wave functions,
rather than the 2P-1S case referred to above. In the analog QCD case,
the quark color magnetic spin flip also does not alter the quark
spatial wave function. This allows us to ignore questions regarding the
interaction of a quark and antiquark in a color octet state and to
proceed with our analysis in the absence of a more analogous, but still
calculable, concept.

Finally, we note that, in the free atomic case, the 2P and 1S states
have differing energies, but the hybrid 1S plus photon state is
necessarily degenerate, in leading order, with the 2P state. In order
to match the 2P-1S energy difference in our atomic magnetic spin flip
model, we include the hyperfine interaction between the electron and
(fixed polarization) proton to split the energies of the two spin
orientation states of the electron. The hybrid state of the lower
energy electron spin orientation state plus the photon is again
necessarily degenerate, in leading order, with the higher energy
electron spin orientation state, in the free atomic case.

However, in both the QCD and cavity atom cases, the hybrid and the
excited states need not be degenerate: In the cavity atom case,
this is due to the energy eigenstates of the photon being determined
by the cavity size, while in QCD we simply don't know which state is
more massive. We therefore take advantage of the flexibility of our
cavity atom, magnetic spin flip model to examine all three cases:
the excited state more massive than the hybrid, less massive, and
approximately degenerate. Of course, when the `excited' state is
less massive, it is actually the lower or ground state, and the
terminology inherited from the free atomic case no longer applies.

\section{Hydrogen in a cavity}

We begin by studying the lowest states of an electron coupled
to a (spin-up) proton.

The interaction of an electron with an electromagnetic field is
described by the Pauli equation:
\begin{eqnarray}
i \hbar { \partial \over{\partial t}}\phi =
 \left[ {1\over{2m}} \left( \vec{p} -{e\over{c}}\vec{A} \right)^2 +e
 A^0 -{e\over{2 m c}} \vec{\sigma }\cdot (\vec{\nabla} \times \vec{A}
 ) \right] \phi 
\label{pauli}
\end{eqnarray}
(This also shows how the electron interacts with the transverse photon.)

In the simplest treatment of the hydrogen atom, the field $A^0$ is
treated as an external field due to an infinitely heavy charge, i.e.,
the proton.  Corrections due to the mass of the proton being finite 
are included in bound state perturbation theory. The field due to a 
single infinitely heavy charge is easily found from Gauss' law. The 
problem then becomes how to satisfy the boundary condition on the 
surface of the cavity, namely, a {\it continuous} and {\it vanishing}
normal component of the electric field.
Contrary to an ordinary, electric superconductor ( in which
the  role of electic and magnetic fields is reversed compared to the
case at hand), a surface charge to neutralize the field of the proton
cannot form.\footnote{ Interestingly, for an electric superconductor
neutralization is achievable even in the presence of a fractional
charge, by a coherent superposition of different net-integer-charge 
surface charge distributions. Violation of the superselection rules 
between differing charge sectors is a consequence (frequently unnoted) 
of the formation of the charged Bose-Einstein condensate of Cooper 
pairs.  Indeed, from the point of view of the pairs, a unit electric 
charge is already a fraction, namely 1/2, of the `unit', $2e$, for the 
effective degrees of freedom in the medium, that is, the Cooper pairs.} 
Therefore, a single charge in a cavity inside a magnetic superconductor 
cannot exist. This is the  appealing physical picture of confinement 
behind the bag model\cite{bagschool}.

Imagine, however, a proton at the center of cavity surrounded by a
unit net charge bound to a spherical surface infinitesimally close 
to the surface. By Gauss' law, this would not affect the electric 
field distribution in the interior of the cavity. We can achieve 
the same effect by using the electron to neutralize the charge of 
the proton, provided the boundary condition becomes the constraint 
that the wavefunction of the atomic electron must vanish at (within 
a penetration length of) the surface of the vacuum bubble. With 
both charges now completely confined to the interior of the vacuum 
bubble, the net total charge presented at the surface is zero and 
so no questions regarding electric flux tubes need be addressed. 
Thus, we can still use the usual Coulomb potential due to the proton 
to describe the atomic electric field, experienced by the electron, 
to the edge of the cavity.

We ignore the issue of the multipole moments of the atom. For
self-consistency in this, the hydrogen atom is a better choice
of a system than positronium, because the proton is better
localized. For positronium, due to motion of the positron, the
Coulomb interaction can be expected to be more affected by the
presence of the boundary, requiring higher multipole moments.
Hence, we choose the hydrogen atom, even though the hybrid
states in QCD are likely to be more interesting (not to mention
confusing!) when the quarks are of comparable mass rather than
in heavy-light systems. For this reason, and to make our model
as simple and intuitive as possible, we choose to consider the
ground state doublet of the hydrogen atom. If the mass of the
proton is sufficiently large, these two states constitute a 
simple two level system.  One would be hard-pressed to find a 
simpler model for hybrid-ordinary meson mixing.

\subsection{Electron wave functions}

We procceed with the description of the atom in a standard way:
First we solve the Schrodinger equation for the electron moving
in the Coulomb potential, and then calculate the hyperfine
splitting in bound state perturbation theory.  For the leading
order, we solve
\begin{eqnarray}
\left( -{1\over{2m}} {\nabla}^2 - {\alpha \over{r}} \right) 
\psi = E_0  \psi \  .
\end{eqnarray}

The general solution for the ground state is the standard,
exponentially suppressed confluent hypergeometric function
(with the normalization, $N^{-1}$, described below)
\begin{eqnarray}
\psi (r) = N^{-1} e^{-{x\over{n_0}} } {_1F _1} (1-n_0, 2,
 {2x\over{n_0}})
\label{wvfcn}
\end{eqnarray}
where we have introduced a dimensionless separation, $x$, and
 eigenvalue,
$n_0$:
\begin{eqnarray}
x & \equiv &  \alpha m r,  \\
 n_0 & \equiv & \sqrt{(\alpha^2 m)/(-2E_0)} .
\end{eqnarray}
We use R to denote the radius of the spherical cavity and $x_0$ to
denote its dimensionless size,
\begin{eqnarray}
x_0 \equiv \alpha m R \  .
\end{eqnarray}

The eigenvalue is determined by the boundary condition
\begin{eqnarray}
\psi (r=R) = 0
\end{eqnarray}
appropriate to this nonrelativistic analysis.\footnote{This 
should be compared with the relativistic condition found in 
the MIT bag model; see, e.g., Ref.\cite{bagschool}. Changing to 
that boundary condition affords an excellent opportunity to 
study relativistic corrections.} We solve this transcendental 
equation for the eigenvalue, $n_0$, numerically. The results 
are shown in the Table. Note that the numerical solution shows 
that, for $x_0 > 3 $, the difference between the eigenvalue in 
the cavity and the continuum limit is already $\leq~6$\%. This 
should not be surprising. The size of the atom is determined 
dynamically, and once the cavity becomes larger than the 
typical Bohr radius, the dynamical structure of the atom is 
not significantly affected.

The normalization of the wavefunction is required to be unity. Hence,
from Eq.(\ref{wvfcn}),
\begin{eqnarray}
N^{-1} & = & 
 \left[ 4 \pi R^3 \int_0^1 dt \ t^2  e^{-2 t {x_0\over{n_0}} } \vert
 _{1}F_{1}(1-n_0, 2, t{2x_0\over{n_0}})\vert ^2 \right]^{-1/2} \  .
\end{eqnarray}

As stated above, we consider the hydrogen ground state doublet.
By tuning the `proton' mass, we can arrange for this doublet to
be arbitrarily separated from the rest of the spectrum, i.e.,
$\Delta E_{spin} \ll \Delta E_{radial, \ orbital}.$ The spin-spin
splitting is due to the standard magnetic (Breit-Fermi) hyperfine
interaction,
\begin{eqnarray}
V_{spin-spin} & = &  {8 \pi \alpha \over{3 m M}}
\delta^3(\vec{r}) \vec{s}_{proton} \cdot \vec{s}.
\end{eqnarray}
associated with the ordinary Coulomb potential.

Let the spin of the proton be fixed in the $+z$ direction. Then
the changes to the electron energies in the ground state doublet
arising from hyperfine interaction are
\begin{eqnarray}
{8 \pi \alpha \over{3 m M}} \vert \psi(0) \vert ^2 {1\over{2}}
\langle s_z \rangle .
\end{eqnarray}
where the $1/2$ in front of the expectation value of $s_z$ comes from
the proton spin and $\langle s_z \rangle = \pm 1/2$ is the expectation
value of the electron spin.  The factor $\vert \psi(0) \vert ^2$ is
just the calculated norm, $N^{-2}$.  (See the Table.)

In the limit of large $R$ (in practice $x_0 > 3 $), the eigenvalues and
the wavefunctions are well approximated by their respective continuum
limits.  The transition between the large $R$ and small $R$ regimes is
relatively sharp, as is evident from the Table.

\subsection{Constituent photon}

The lowest mode in the cavity with magnetic boundary conditions is
\cite{trn}
\begin{eqnarray} 
\vec{A} = N_{\gamma}^{-1} j_1(\omega_0 r)
\vec{L}Y_{1m}(\vec{\Omega}) a_m + H.C.  \label{vecA}
\end{eqnarray}
where $a_m$ is the mode annihilation operator, $j_1$ is the spherical
Bessel function, $\vec{L}$ is the angular momentum operator, the
$Y_{lm}$ are the spherical harmonics and $\vec{r} = r\vec{\Omega}$. 
The frequency $\omega_0 = k_0/R$ is fixed by the boundary condition:
\begin{eqnarray}
k_0 j_1'(k_0) = -j_1(k_0)
\end{eqnarray}
which is determined by the requirement that the photon flux leaving
the cavity vanishes. The solution is $k_0 = 2.7437$~\cite{trn}. The
normalization of this wavefunction to correspond to a single photon
is given by $N_{\gamma}$,
\begin{eqnarray}
N_{\gamma}^{-1}& = &{1\over{\sqrt{2 \omega_0}}}\left[ 2\int_0^{R} dr \
 r^2
j_1^2(\omega_0 r)
\right]^{-1/2} \nonumber\\
& = & R^{-1} \left[ 2 k_0 \left({1\over{k_0^2} }
+ {\sin (k_0) \cos (k_0)\over{k_0^3}}
-2 {\sin^2 (k_0) \over{k_0^4}} \right) \right]^{-1/2} .
\end{eqnarray}
The term in parenthesis in the last expression is a number,
$$c(k_0) = 1.2857,$$
which is independent of $R$, i.e., $N_{\gamma}^{-1} = c(k_0)R^{-1}$.
Note that, since $x_{0} = \alpha m R$,
\begin{eqnarray}
\omega_{0} = \alpha m \frac{k_{0}}{x_{0}}
\end{eqnarray}
so that the energy of this lowest photon mode is $k_{0}/{x_{0}}$ in
units of $\alpha m$.

\subsection{Emission of the constituent photon.}

{}From the Pauli equation, Eq.(\ref{pauli}), the operator for the
interaction of the electron with the constituent photon is:
\begin{eqnarray}
 (-){e\over{2 m c}} \vec{\sigma }\cdot (\vec{\nabla} \times \vec{A} ).
 \label{op1}
\end{eqnarray}

First, we find the  magnetic field associated with the constituent
photon from Eq.(\ref{vecA}).
\begin{eqnarray}
\vec{\nabla}  \times \vec{A}_m  & = &
 i \sqrt{{3\over{4\pi}}} \left[ \hat{e}_r 2 {j_1\over{r}} \cos \theta
 - \hat{e}_{\theta}\left({\partial j_1\over{\partial r}} + {j_1\over{r}}
 \right)
 \sin \theta \right], \ \   m=0 \nonumber\\
  =  - i& \sqrt{{3\over{8\pi}}}& \left[ \hat{e}_r 2 {j_1\over{r}} \sin
 \theta
e^{i \phi}
 + \left({\partial j_1\over{\partial r}} + {j_1\over{r}} \right)
 \left( \hat{e}_{\theta}\cos \theta e^{i \phi}+ \hat{e}_{\phi}i e^{i
\phi}\right) \right], \ \   m=+1 \nonumber\\
 \vec{\nabla} \times \vec{A}_{-m } & = &(\vec{\nabla} 
 \times \vec{A}_m)^{\ast}   \  .
\end{eqnarray}

We need only the spin-flip part of the interaction
\begin{eqnarray}
\vec{\sigma} \cdot \vec{B} = \sigma_+ B_- + \sigma_- B_+  + \sigma_z B_z
\end{eqnarray}
where
$B_{\pm} = (B_x \pm i B_y)/{\sqrt{2}}= (\hat{x}\pm i \hat{y}) \cdot
(\vec{\nabla} \times \vec{A})/{\sqrt{2}}. $

Using the expression for specific $m$ values, it  is straightforward to
show that
\begin{eqnarray}
m=0:  &  \sigma_{\stackrel{-}{+} } B_{\stackrel{+}{-} } \propto  e^{ \pm
 i \phi}
\nonumber\\
m=+1: &     \sigma_{\stackrel{-}{+} } B_{\stackrel{+}{-} } \propto
 e^{+i \phi
\pm i\phi}
\nonumber\\
m=-1: &     \sigma_{\stackrel{-}{+} } B_{\stackrel{+}{-} } 
\propto e^{-i \phi \pm i\phi}  \  .
\end{eqnarray}
The ground state wavefunction is spherically symmetric, therefore the
$m=0$ mode does not contribute at all. The $m=1$ photon gives a nonzero
expectation value for $\sigma_+ B_-$; $m=-1$ similarly gives a value
for $\sigma_- B_+$ and is the complex conjugate of the $m=1$ case. In
what follows, we concentrate on the mixing between a spin up pure
electron state ($\vert a \rangle \equiv \vert \uparrow \rangle$), and
a spin down electron with a constituent photon ($\vert b \rangle \equiv
\vert \downarrow \gamma \rangle$). Therefore, we need only
the $m=+1$ case.

In the expectation value of the operator in Eq.(\ref{op1}), the
spin and spatial expectation values factor out:
\begin{eqnarray}
\langle a \vert {{-e}\over{2 m c}} \sigma _+ B_- \vert b \rangle =
(-){e\over{2 m c}} \langle  \sigma _+ \rangle \, I
\end{eqnarray}
where
\begin{eqnarray}
\lefteqn{I \equiv  -{i\over{4}} \sqrt{3\over{\pi}} N^{-2}
 N_{\gamma}^{-1}
\int  d^3 r \, e^{-{2 \alpha m r \over{n_0}} } \vert
_{1}F_{1}(1-n_0, 2, {2 \alpha m r \over{n_0}})\vert^2} \nonumber\\
& & \times \left[2 {j_1(\omega_0 r)\over{r}}(1- \cos^2 \theta )
 + \left({\partial j_1(\omega_0 r)\over{\partial r}} +
 {j_1(\omega_0 r)\over{r}} \right)
 \left( 1 + \cos^2 \theta \right) \right] .
\end{eqnarray}
After straightforward algebra, the integral reduces to
\begin{eqnarray}
I = -i \sqrt{3\over{\pi}} N^{-2} N_{\gamma}^{-1} 
{4 \pi \over{3}} \int dr \ r \sin (\omega_0 r) 
e^{-{2 \alpha m r \over{n_0}} } \vert _{1}F_{1}(1-n_0, 2, 
{2 \alpha m r \over{n_0}})\vert^2 \  . \label{I}
\end{eqnarray}

\subsection{ Special case: $R$ large. }

For the sake of consistency, $\alpha$ in our calculation has to be
small to justify the nonrelativistic treatment of the hydrogen atom.
In addition to this restriction, the ratio $m/M$ has to be small so
that the proton is effectively static and the spherical cavity is a
good description. Together, these imply that the energy shifts of
the doublet must be small compared to $\alpha m$.

The energy of the photon, $k_0\over x_0$, as measured in units of
$\alpha m$, is of order one for $x_0 = 3$. (Recall that the large
$R$ regime starts around $x_0 \simeq 3$). This is large, not only
compared to the spin-spin interaction energy, but  also compared to
the radial binding, which is of order $\alpha$ in these units.
Thus, it is sufficient to consider just the large $R$ limit, i.e.,
$x_0 \gg 3$, to obtain results over a range of photon energies
from large to small relative to all of the energy scales of pure
hydrogen.

In the large $R$ limit, the radial part of the wavefunction, $\psi(r)$,
and its value at the origin, $N^{-1}$, are well approximated by their
respective continuum values, i.e.
\begin{eqnarray}
n_0 & \doteq & 1 \\
\psi(r) & \doteq &  N^{-1} e^{-x} \\
N^{-1} & \doteq  & \left[  \pi (\alpha m)^{-3} \right]^{-1/2} .
\end{eqnarray}

The energies of the two states under consideration, i.e. the spin
up electron (spin one hydrogen state) and the spin down electron
(including a spin zero hydrogen component) together with a
constituent photon (for a total spin of one, again) are
\begin{eqnarray}
E_a  \equiv E(\uparrow) & =& \alpha m  \  \left( {2\over{3}} \,
{m\over{M}} \alpha^3 \right) \\
E_b  \equiv  E(\downarrow +\gamma) &=& \alpha m  \  \left( -{2\over{3}}
 \,
{m\over{M}} \alpha^3 + {k_0\over{x_0}} \right),
\end{eqnarray}
relative to the centroid of the doublet.

Since the wavefunction reduces to a simple exponential, the integral
in Eq.(\ref{I}) can be evaluated analytically,
\begin{eqnarray}
&I &=  - i  {16\over{\sqrt{3 \pi}}}  c(k_0) k_0
 { x_0^2  \over{ \left(4 x_0^2 +k_0^2\right)^2}}
 (\alpha m)^2  \ .
\end{eqnarray}

The full transition energy matrix element between the two states is
 then:
\begin{eqnarray}
T_{ab} \equiv \langle a \vert H_I \vert b \rangle
& = &i \, \alpha m \, \sqrt{2\over{3}} {c(k_0)\over{k_0}} {\left(
{k_0\over{x_0}}\right)^2
\over{\left[ 1 + \left( {k_0\over{2 x_0}}\right)^2\right]^2}} \,
\alpha^{3/2} \ .
\end{eqnarray}

The Hamiltonian of this two-level system is therefore:
\begin{eqnarray}
H=\left(  \begin{array}{cc}
E_a & T_{ab} \\
T_{ab}^* & E_b
\end{array}  \right) \ \ ,
\end{eqnarray}
and its eigenstates are
\begin{eqnarray}
N_1 \vert 1 \rangle & = & |T_{ab}| \, \vert a \rangle
   + \left( E_1 - E_a \right) \, \vert b \rangle   \\
N_2 \vert 2 \rangle & = & |T_{ab}| \, \vert a \rangle
   + \left( E_2 - E_a \right) \, \vert b \rangle
   \label{eigenstates}
\end{eqnarray}
where
$N_{1,2} = [|T_{ab}|^2 + \left( E_{1,2} - E_a \right)^2]^{1/2}$
($E_a$, $E_b$, and $T_{ab}$ are as given in previous expressions),
and where we have absorbed a phase of $\pm i$ into the definition
of (either) one of the states for convenience. The corresponding
eigenvalues are
\begin{eqnarray}
E_{1,2} = {\alpha m \over{2}} \left[{k_0\over{x_0}} \pm
\sqrt{ \Delta^2 + {2\over{3}}\left({{2c(k_0)}\over{k_0}}\right)^2
 {\left({k_0\over{x_0}}\right)^4\over{\left[ 1
 + \left( {k_0\over{2 x_0}}\right)^2\right]^4}} \,
 \alpha^3}\right] 			\label{eiegenE}
\end{eqnarray}
where
\begin{eqnarray}
\Delta \equiv {4\over{3}}{m\over{M}} \alpha^3 -{k_0\over{x_0}}. \label{DELT}
\end{eqnarray}
is the splitting between the $(a,b)$ states before mixing, in units
of $\alpha m$. The energy difference between the eigenstates is
\begin{eqnarray}
{{\delta E}\over{\alpha m}} & \equiv &
     {(E_1 - E_2)\over{\alpha m}} \nonumber  \\
& = & \sqrt{\Delta^2 + {2\over{3}}\left( {2 c(k_0)\over{k_0}}\right)^2
 {\left({k_0\over{x_0}}\right)^4\over{\left[ 1 +
 \left( {k_0\over{2 x_0}}\right)^2\right]^4}} \, \alpha^3} \label{ediff}
\end{eqnarray}
(All energies in this regime are much less than $\alpha m$ for
sufficiently large values of $x_{0}$.) Since, for the values of
$k_0$ and $c(k_0)$ the numerical value of $ \left( 
{2 c(k_0)\over{k_0}}\right)^2 \sim 1$, in what follows, we omit
this combination in approximate formulas below, for simplicity.

\section{Discussion}

Below, we will refer to the pure hydrogenic (proton plus electron only)
states, corresponding to quark-antiquark ("pure quark") states in QCD, 
as pure electron states. The states including a cavity-trapped photon,
corresponding to hybrid quark-antiquark-gluon states in QCD, will be
referred to as `hybrid' states. There are three distinct limits defined
by the relative size of the unperturbed energies:

\subsection{Photon energy much less than the energy splitting between
the pure electron states}

In this case, the upper pure electron state corresponds to the QCD 
case of a quark-antiquark color singlet excitation of overall higher 
energy, whereas the hybrid state (lower pure electron state plus 
photon) corresponds to a lower energy quark-antiquark color octet 
state plus gluon making up a hybrid state of overall energy which 
is still lower than that of the excited singlet. We have
\begin{eqnarray}
{k_0\over{x_0}} <<  {4\over{3}}{m\over{M}} \alpha^3. \label{lowgam}
\end{eqnarray}
This occurs when the cavity becomes so large that effectively the
energy of the lowest lying photon approaches the continuum limit,
i.e. zero.

Of course, as the cavity size increases, the energy of the excited
photons decreases also, and eventually some photon state would become
comparable with the energy splitting of the hyperfine states. We are
not interested in this scenario, since we are not concerned with the
dynamics of this system for itself. Rather, we use the size of the
cavity to model the magnitude of the confining energy of the gluon in a
hybrid state. From this point of view, the case at hand corresponds to
an ordinary meson and a hybrid  which contains a constituent gluon of
energy much smaller than the energy difference of the pure
quark-antiquark components. Such a case may occur in QCD if the next
higher gluonic state turns out to be very much higher in energy, or if
some quantum number constraint prevents mixing between the pure
quark-antiquark components and a gluonic state of energy closer to the
energy difference of the pure quark-antiquark components, corresponding
in our model to one of the higher cavity photon states.

The energies of the eigenstates are, respectively,
\begin{eqnarray}
E_{1,2}/(\alpha m) = \pm {2\over{3}}{m\over{M}} \alpha^3 + {1\over{2}}
\left( {k_0\over{x_0}}
\stackrel{-}{+}{k_0\over{x_0}} \right).
\end{eqnarray}
The energy difference between the eigenstates is then
\begin{eqnarray}
{{\delta E}\over{\alpha m}} = {4\over{3}}{m\over{M}} \alpha^3
	-{k_0\over{x_0}} = \Delta. \label{DEL}
\end{eqnarray}

Not surprisingly, the mixing is small, as $E_{1}-E_{a}$ is of the
order of the fourth power of the photon energy:
\begin{eqnarray}
\vert 1 \rangle & \simeq &  \vert a \rangle
	+{\cal O}\left[\sqrt{{3\over{2\alpha^3}}} {M\over{4m}}
	\left({k_0\over{x_0}}\right)^2\right] \vert b \rangle\\
\vert 2 \rangle & \simeq &  \, \vert b \rangle
	-{\cal O}\left[\sqrt{{3\over{2\alpha^3}}} {M\over{4m}}
	\left({k_0\over{x_0}}\right)^2\right] \vert a \rangle.
\end{eqnarray}
Despite the inverse powers of $\alpha$ and the ratio ${M\over{m}}$, 
the corrections from the mixing are extremely small in this case, 
in view of Eq.(\ref{lowgam}). The hybrid state has the lower energy 
of the two and, again in view of Eq.(\ref{lowgam}), $\Delta \approx 
{4\over{3}}{m\over{M}} \alpha^3$ so the state has almost the same 
energy as the lower pure electron state.

\subsection{Photon energy comparable to the energy splitting of the
pure electron states}

This corresponds to the QCD case where the quark-antiquark color
singlet excited state is comparable in overall energy to the hybrid
state.

With decreasing size of the cavity, the energy of the constituent
photon increases and it becomes comparable to the energy
splitting between the pure electron states. With $\Delta$ as 
defined above in Eq.(\ref{DELT}), we have
\begin{eqnarray}
 | \Delta |<< {k_0\over{x_0}} \sim \  \  
{4\over{3}}{m\over{M}} \alpha^3 \  . \label{small}
\end{eqnarray}
As long as $\Delta$ is larger than the magnitude of the off diagonal
matrix element of the Hamiltonian, the system remains similar to the
situation described in the previous section. However, when the energy
splitting $\Delta$ is also much smaller than the mixing energy, that 
is,
\begin{eqnarray}
  {\Delta}^2 << \left({k_0\over{x_0}}\right)^4 \alpha^3 \  ,
\label{smaller} 
\end{eqnarray}
the energies of the eigenstates are approximately
\begin{eqnarray} E_{1,2}/(\alpha m)
= {1\over{2}} \left\{ {k_0\over{x_0}} \pm \left({k_0\over{x_0}}\right)^2
\sqrt{{2\over{3}} \alpha^3} \pm {\Delta^2
\over{2\left({k_0\over{x_0}}\right)^2 \sqrt{{2\over{3}} \alpha^3}}}
\right\}. 
\end{eqnarray}
(Since ${k_0\over{x_0}} \ll 1$ here, to satisfy Eq.(\ref{small}), 
the term in square brackets in the denominator of Eq.(\ref{ediff}) 
is approximately unity.) The wave functions are, to ${\cal O}(\Delta)$,
\begin{eqnarray}
\vert 1 \rangle & = & {1\over{\sqrt{2}}} \left\{
\left[ 1-{\Delta \over{2 \left({k_0\over{x_0}}\right)^2 \sqrt{{2\over{3}} 
\alpha^3}}} \right] \vert a \rangle +
\left[ 1+{\Delta \over{2 \left({k_0\over{x_0}}\right)^2 \sqrt{{2\over{3}} 
\alpha^3}}} \right] \vert b \rangle \right\}  \\
\vert 2 \rangle & = & {1\over{\sqrt{2}}} \left\{
\left[ 1+{\Delta \over{2 \left({k_0\over{x_0}}\right)^2 \sqrt{{2\over{3}} 
\alpha^3}}} \right] \vert a \rangle +
\left[ -1+{\Delta \over{2 \left({k_0\over{x_0}}\right)^2 \sqrt{{2\over{3}} 
\alpha^3}}} \right] \vert b \rangle \right\}  . 
\end{eqnarray}
The mixing is maximum when $\Delta=0$, as one would expect. 
(The presence of $\alpha$ in the denominator may be misleading:  
Eq.(\ref{smaller}) shows that, in this case, $|\Delta| << 
\alpha^{\frac{3}{2}}$, as ${k_0\over{x_0}} \ll 1$ in the large 
$R$ limit considered in this paper.  So, in fact, the term with 
the inverse power of the coupling constant is nonetheless very 
small.)

The energy splitting between physical states is, not surprisingly,
dominated by the off diagonal matrix element of the Hamiltonian, viz.
\begin{eqnarray} 
{\delta E}\over{\alpha m} &=& \left({k_0\over{x_0}}\right)^2 
\sqrt{{2\over{3}} \alpha^3} \left[ 1 + {{3 \Delta^2} \over 
{ 4 \alpha^3}} \left({x_0\over{k_0}}\right)^4 \right] .
\end{eqnarray}
(Again, despite the misleading inverse powers of $\alpha$, the
correction term in the square brackets is small due to
Eq.(\ref{smaller}).)

\subsection{Photon energy much larger that the energy splitting of the
pure electron states}

This corresponds to the QCD case where the hybrid state has much larger 
overall energy than the quark-antiquark color singlet excited state.

As the cavity size is decreased further, the photon energy increases
as $x_0^{-1}$ and can become much larger than the energy splitting
of the pure electron states. We can arrange  for both
${4\over{3}}{m\over{M}}\alpha^3$ and ${k_0\over{x_0}}$ to be much less
than one in the $R$ large limit, while satisfying
\begin{eqnarray}
{4\over{3}}{m\over{M}}\alpha^3<<{k_0\over{x_0}}. 
\end{eqnarray}
Thus, $\Delta \approx -{k_0\over{x_0}}$ and the eigenvalues are
\begin{eqnarray}
E_1 /\left(\alpha m \right) & = &  
{k_0\over{x_0}} \left[ 1+ \left({k_0\over{x_0}}\right)^2{\alpha^3\over{6}} 
\right] - {2\over{3}}{m\over{M}}\alpha^3 \\
E_2 /\left(\alpha m \right) & = & {2\over{3}}{m\over{M}}\alpha^3
-\left({k_0\over{x_0}}\right)^3{\alpha^3\over{6}}  \  .
\end{eqnarray}
Hence, we have that
\begin{eqnarray}
{{\delta E}\over{\alpha m}} &\approx& -\Delta +
\left({k_0\over{x_0}}\right)^{3}{\alpha^3\over{3}} \\
&=& {k_0\over{x_0}} - {4\over{3}}{m\over{M}}\alpha^3
+\left({k_0\over{x_0}}\right)^{3}{\alpha^3\over{3}}.
\end{eqnarray}
However, since ${k_0\over{x_0}} < 1$, in this case the difference is
almost the same as the energy of the trapped cavity photon,
corresponding to the constituent gluon in the QCD hybrid case.

The wave functions in this case are
\begin{eqnarray}
\vert 1 \rangle & = &  \sqrt{{\alpha^{3}}\over{6}} \left( 
{k_0\over{x_0}}\right)  \vert a\rangle  + \left[ 1-
{{\alpha^{3}}\over{12}}\left( {k_0\over{x_0}}\right)^2 
\right] \vert b \rangle \\
\vert 2 \rangle & = & 
\left[ 1- {{\alpha^{3}}\over{12}}\left( {k_0\over{x_0}}\right)^2
\right] \vert a \rangle  -\sqrt{{\alpha^{3}}\over{6}} \left(
{k_0\over{x_0}}\right) \vert b \rangle \  .
\end{eqnarray}
Note that the energy of the higher state, $E_1$, remains
dominated by the diagonal element, but in this case, it
is the hybrid state.  The correction from mixing is further
suppressed by $\alpha^3$. Conversely, mixing becomes a crucial 
factor for the energy of the lower state when ${k_0\over{x_0}}$ 
is comparable to $\left[{m\over{M}}\right]^{1/3}$. The energy 
of the lower state, $E_2$, is affected by the mixing even 
though the state is dominated by the pure non-hybrid component. 
In particular, cancellation or near-cancellation between the 
spin-splitting and mixing energy terms may lead to the appearance 
that this pure electron (non-hybrid) state is at the wrong energy 
to be identified as a conventional (meson) state within a framework 
that ignores constituent bosons, e.g. a quark model.

\section*{Concluding remarks}

We have examined the physical system of hydrogen in a cavity
and found that, even in the large cavity, small coupling,
heavy nucleus limit, the system has a rich range of available
characteristics which may illuminate corresponding cases of
QCD hybrids, where none of the limits apply. (All parameter
ratios are near unity.)

We find that the state with the constituent boson always mixes
with the state of pure constituent fermions, but if the energy
of the constituent boson itself is much smaller than the energy
difference of the pure fermionic part, the mixing becomes
negligibly small.  Similarly, when the energy of the constituent
boson is much larger than that energy difference, the more
energetic state is dominated by the component with the extra
boson. However, the energy of the lower lying state may or may
not be given by the the state nearly purely composed of fermions,
depending on the relative size of the boson and fermion state
splitting energies.

When the energies of states prior to mixing are comparable, the
mixing is maximum when the energy of the boson coincides with
the size of the energy difference of the fermionic part.

Perhaps the most interesting and most relevant case for our
understanding of the role of hybrid states in QCD is that when
the energy of the constituent boson is larger than the energy
difference of the fermionic part. Our results suggest that, 
if the mixing of the hybrid with the pure fermion states is
strong enough in this case, then the energy of the nominally
pure quark-antiquark color singlet state is significantly
altered even though the wavefunction of the state is essentially
unaffected.  Note that this may well occur in QCD, where
${k_0\over{x_0}} \sim \alpha_{S} \sim {m\over{M}} \sim 1$. This
may have important implications for analyses of decay patterns
that can influence the interpretation of states as quark-antiquark
or hybrid. In particular, the state may lack evidence of hybrid
decay patterns even though its mass suggests that it does not
belong in a representation of pure quark-antiquark states, and
thus superficially requires an alternate interpretation.

Continuation to other parameter regimes of interest ($\alpha
\sim 1, {m\over{M}} \sim 1$) may appear to be straightforward
using numerical methods. Unfortunately, the approximations on which
the model is based become invalid, in general, which would
require more involved analysis to remedy. However, it should
be fairly straightforward to study this problem with a cavity of
arbitrary size or relativistic boundary conditions.

As lagniappe, we note that, in addition to the theoretical certainty of 
the calculations of such a well understood system presented here, it may 
be possible to realize such systems experimentally using magneto-optical 
traps (MOTs) for alkali atoms. By adjusting the magnetic field and choice 
of states, it may be possible to realize a spin flip system with a long 
enough wavelength that an exterior conducting sphere could be added and 
still allow for the entry of laser beams required for the MOT. This would 
allow for the experimental study of electron-photon hybrid states as a 
model for QCD hybrids. 

\section*{Acknowledgments}
We would like to thank Charles Thorn and Philip Page for
discussions.  This research is supported by the Department
of Energy under contracts W-7405-ENG-36 and DE-FG02-87ER40365.

\begin{table}
\caption{Numerical results for the ground state hydrogen atom and
ground state photon.}
{
\begin{tabular}{|c|c|c|c|} \hline
   $x_0$ & $n_0 $ & $N^{-2} $ in units $(\alpha m )^{3}$ & $k_0/x_0$
\\ \hline
\hline
$1.8456$ & $7.38261$ & $0.8653$ & $1.48662$ \\ \hline
$1.8768$ & $3.75359$ & $0.840668$ & $1.4619$ \\ \hline
$2.0$ & $2.0$  & $0.755681$ & $1.37185$ \\ \hline
$2.03412$ & $1.84$  & $0.733805$ & $1.34884$ \\ \hline
 $2.200$ & $1.467$ & $0.6505$   & $1.24714$
 \\ \hline
 $2.4712$ & $1.23562$ & $0.5519$ & $1.11027$
    \\ \hline
        $3.1876 $ & $1.0625$ & $0.4161 $ & $ 0.86842$  \\ \hline
  $4.06228$ & $1.01557$ & $0.352562$  & $0.675409$  \\ \hline
 $5.0175$ & $1.0035$ & $0.3285$  & $0.546826$
    \\ \hline
 $6.004 $ & $1.00072$ & $0.32094$  & $0.456979$    \\ \hline
 $\infty$ & $1.0$ & $0.31831$  & $0$   \\ \hline
\end{tabular}
}
\vskip0.2in%
\end{table}

\end{document}